\documentclass[aip,jap,reprint]{revtex4-1}
\usepackage{graphicx}
\usepackage{ulem}
\usepackage{amssymb}
\usepackage{amsmath}
\usepackage{wasysym}
\usepackage{relsize}
\begin{document}

\newcommand{\be}{\begin{equation}}
\newcommand{\ee}{\end{equation}}
\newcommand{\bea}{\begin{eqnarray}}
\newcommand{\eea}{\end{eqnarray}}
\newcommand{\Tbar}{{\bar{T}}}
\newcommand{\En}{{\cal E}}
\newcommand{\K}{{\cal K}}
\newcommand{\U}{{\cal U}}
\newcommand{\GC}{{\cal \tt G}}
\newcommand{\Lop}{{\cal L}}
\newcommand{\DB}[1]{\marginpar{\footnotesize DB: #1}}
\newcommand{\q}{\vec{q}}
\newcommand{\kt}{\tilde{k}}
\newcommand{\Lopn}{\tilde{\Lop}}
\newcommand{\noi}{\noindent}
\newcommand{\ovn}{\bar{n}}
\newcommand{\ovx}{\bar{x}}
\newcommand{\ovE}{\bar{E}}
\newcommand{\ovV}{\bar{V}}
\newcommand{\ovU}{\bar{U}}
\newcommand{\ovJ}{\bar{J}}
\newcommand{\calE}{{\cal E}}
\newcommand{\calF}{{\cal F}}
\newcommand{\calA}{{\cal A}}
\newcommand{\calC}{{\cal C}}
\newcommand{\ovphi}{\bar{\phi}}
\newcommand{\zt}{\tilde{z}}
\newcommand{\ttl}{\tilde{\theta}}
\newcommand{\nuv}{\rm v}
\newcommand{\ds}{\Delta s}
\newcommand{\fn}{{\small {\rm  FN}}}
\newcommand{\cc}{{\cal C}}
\newcommand{\cd}{{\cal D}}
\newcommand{\tth}{\tilde{\theta}}
\newcommand{\ttr}{\tilde{\rho}}
\newcommand{\cb}{{\cal B}}
\newcommand{\cg}{{\cal G}}
\newcommand{\cf}{{\cal F}}
\newcommand{\tr}{\text{tr}}
\newcommand{\inj}{\text{in}}
\newcommand{\SCL}{\text{SCL}}
\newcommand{\CL}{\text{CL}}
\newcommand\norm[1]{\left\lVert#1\right\rVert}

\title{Predicting space-charge affected field emission current from curved tips}

\vskip 0.15 in

\author{Debabrata Biswas}\email{dbiswas@barc.gov.in}
\affiliation{
Bhabha Atomic Research Centre,
Mumbai 400 085, INDIA}
\affiliation{Homi Bhabha National Institute, Mumbai 400 094, INDIA}
\author{Raghwendra Kumar}
\affiliation{
Bhabha Atomic Research Centre,
Mumbai 400 085, INDIA}
\author{Gaurav Singh}
\affiliation{
Bhabha Atomic Research Centre,
Mumbai 400 085, INDIA}
\affiliation{Homi Bhabha National Institute, Mumbai 400 094, INDIA}


\begin{abstract}
  Field emission studies incorporating the effect of space charge reveal that for planar
  emitters, the steady-state field $E_P$, after initial transients, settles down to a value
  lower than the vacuum field $E_L$. The ratio $\vartheta = E_P/E_L$ is a measure of the severity of
  space charge effect with $\vartheta = 0$ being most severe and $\vartheta \simeq 1$ denoting the
  lack of significant effect. While, $E_L$ can be determined from a single numerical evaluation
  of the Laplace equation, $E_P$ is largely an unknown quantity whose value can be approximately
  found using physical models or can be determined `exactly' by particle-in-cell or
  molecular dynamics codes. We propose here a simple model that applies to planar as well as
  curved emitters based on an application of Gauss's law. The model is then refined using
  simple approximations for the magnitude of the anode field and the spread of the beam when it
  reaches the anode. The predictions are compared with existing molecular dynamics results
  for the planar case and  particle-in-cell simulation results using PASUPAT for curved emitters.
  In both cases, the agreement is good. The method may also be applied to large area field emitters
  if the individual enhancement factors are known, for instance, using the hybrid model
  (D.Biswas, J. Vac. Sci. Technol. B 38, 063201 (2020)).
  
\end{abstract}

\maketitle

\section{The space charge affected current}

\subsection{Introduction}

Field emission refers to the process by which electrons tunnel out from the surface of
a conductor on application of a strong electric field\cite{FN,Nordheim,MG,forbes2006}. The height and width of the
tunneling potential barrier depends on the strength of the electric field. As a consequence, the
field-emission current depends sensitively on the  local electric field on the emitter surface with
a small change in field resulting in a large change in emitted current.

The presence of field-emission electrons in a diode can itself be a cause for change in the
local field on the emitter surface\cite{stern29,ivey49,barbour53,vanVeen,jensen97a,jensen97b,forbes2008}.
If the applied macroscopic field $E_0$ is large enough to cause
sufficient electron emission, the negative charge cloud lowers the magnitude
of the local field on the emitter surface, thereby leading to a decrease in the field emission
current. It may also happen that the local field becomes zero
and emission stops altogether until the
space charge  moves away from the cathode
and is eventually lost from the diode. Thus, the field emission current in a diode
can be oscillatory initially till the local field at the emitter surface saturates
with time and a steady-state prevails \cite{feng2006,jensen2015,torfason2015}.

Attempts to determine\cite{stern29,ivey49,barbour53,forbes2008} the steady-state
cathode electric field and the emission current density for planar emitters
have been made since the basic formulation of field emission by
Fowler and Nordheim\cite{FN} (FN) in 1928 and subsequent corrections and
approximations to the expression for the field emission current density for conductors
\cite{MG,forbes2006}. The connection between space charge and the current density
is established by expressing the charge density $\rho = J_P/v$ where $J_P$ is the
steady-state current density
and $v$ the speed\cite{puri2004}. The speed $v$ can be further expressed in terms of the potential $V$ using
energy conservation. Thus, the Poisson equation can be expressed in 1-dimension as

\be
d^2 V/dz^2 = \kappa J_P/\sqrt{V}  \label{eq:poi}
\ee

\noi
where $\kappa = \epsilon_0^{-1} \sqrt{m/2e}$
where $m$ and $e$ refer to the mass and charge of the electron.
Assuming $J$ to be constant in a parallel-plate diode, Eq.~(\ref{eq:poi}) can
be solved with $V = 0$ at $z = 0$ (grounded cathode) and
$V = V_g$ at the anode placed at $x = D$ to get a relation\cite{barbour53,forbes2008}
between the steady-state field at the cathode $E_P$ and the current density $J_P$:

\be
6\kappa^2J_P^2 D - E_P^3 = (4\kappa J_P V_g^{1/2} + E_P^2)^{1/2}(2\kappa J_P V_g^{1/2} - E_P^2). \label{eq:barbour}
\ee

\noi
This has to be solved self-consistently with a suitable\cite{suitable,forbes2007,forbes2008b,DB_RR}
field emission equation $J_P(E_P)$ in order to determine the space-charge affected field emission current
density $J_P$ in terms of the applied voltage $V_g$. 
It predicts for instance a saturation-like behaviour in the current
density when used with the Murphy-Good\cite{MG} field emission expression for the
current density. In an FN-plot, this gets manifested as a deviation from a straight
line at high cathode fields.

\subsection{space-charge affected current in the planar case}

In a planar situation therefore, space charge does affect the field emission current.
If we denote the
electrostatic field in the absence of any charge by $E_L$
and the saturated field (after field emission has continued for a few transit times)
by $E_P$, a possible measure\cite{alternate,db_scl} of the severity of space charge effect may be
taken to be the field reduction factor\cite{forbes2008}, $\vartheta = E_P/E_L$.
The subscript $L$ and $P$ refer to the Laplace and Poisson equations respectively \cite{forbes2008}.
Clearly $0 \leq \vartheta \leq 1$ with $\vartheta \simeq 1$ denoting the lack of significant
space charge effect and $\vartheta = 0$ being the classical space charge limit \cite{child,langmuir,langmuir23,db_scl}
where $E_P = 0$ and $J_P = J_{\CL}$ where $J_{\CL} = (4/9\kappa) V_g^{3/2}/D^2$ is the Child-Langmuir
current.

The dependence of the space-charge affected current $J_P$ on the
field reduction factor $\vartheta$ can be simplified by re-writing Eq.~(\ref{eq:barbour})
in terms of the dimensionless quantity $\xi = \kappa J_P V_g^{1/2}/E_L^2$ where $V_g/D = E_L$.
The normalization of
$J_P$ in effect is with respect to the planar Child-Langmuir current
and at the space charge limit $\vartheta = 0$, $\xi$ assumes
the value $\xi_{\CL} = 4/9$. Thus, Eq.~(\ref{eq:barbour}) reduces to

\be
6\xi^2 - \vartheta^3 = (4\xi + \vartheta^2)^{1/2}(2\xi - \vartheta^2)
\ee

\noi
which can be further simplified to yield

\be
3\vartheta^2 (1 - \vartheta) = \xi (4 - 9\xi).  \label{eq:forbes1}
\ee

\noi
When $\vartheta = 1$, $\xi$ or $J_P = 0$, while $\vartheta = 0$ corresponds to $\xi = 4/9$.

\subsection{Extension of existing theory for curved emitters}

It is difficult to extend the relation directly to the case of curved emitters since $J_P$
is no longer uniform. However, a plausible phenomenological relation may be
arrived at by recalling a recent result on the space charge limited current for axially
symmetric curved emitters. It states that the space charge limited current
$I_{\SCL} \approx \pi b^2 \gamma_a J_{\CL}$ where $b$ is the radius of the base of
the emitter and $\gamma_a$ is the apex field enhancement factor (a constant)\cite{db_universal}.
Thus, it is natural to define the scaled field and current density as

\bea
\tilde{\vartheta} & = & \frac{E_P}{\gamma_a V_g/D} \label{eq:thet} \\
\tilde{\xi} & = &  \frac{\tilde{\omega} J_P}{\gamma_a J_{CL}} \label{eq:xi} \\
J_P & = & \frac{I_P}{\pi b^2} \label{eq:jp}
\eea

\noi
where $I_P$ is the net space-charge affected emitter current and $\tilde{\omega}$
is a factor that is hitherto unknown. A plausible
relation between the space-charge affected current and apex field can
thus be obtained by substituting $\xi$ and $\vartheta$ in Eq.~(\ref{eq:forbes1})
so that

\be
3\tilde{\vartheta}^2 (1 - \tilde{\vartheta}) = \tilde{\xi} (4 - 9\tilde{\xi}).  \label{eq:curved0}
\ee

\noi
Eq.~(\ref{eq:curved0}) has the correct limiting behaviour for an axially symmetric
curved emitter mounted in a parallel plate diode configuration. The weak space
charge regime has the solution

\be
\tilde{\vartheta} = 1 - \frac{4}{3} \tilde{\xi} + \mathcal{O}(\tilde{\xi})^2 \label{eq:small}
\ee

\noi
which can be equivalently expressed as

\be
\tilde{\vartheta} = 1 - \frac{4}{3} \frac{\tilde{\omega}}{\gamma_a}\xi + \mathcal{O}(\xi)^2 \label{eq:small1}
\ee

\noi
with $\tilde{\omega}$ as a fitting parameter and $\xi$ evaluated with $J_P$ as defined in Eq.~(\ref{eq:jp}).
The linear behaviour predicted by Eq.~(\ref{eq:small1}) has recently been observed
in a PIC simulation\cite{rk2021} where the normalized current
density was defined as $\xi' = I_P/(\pi (gR_a)^2 J_{CL})$ with $g$ taking values between 0.5 and 1.
In the ($\xi',\tilde{\vartheta})$ plane, Eq.~(\ref{eq:small1}) is expressed as

\be
\tilde{\vartheta} = 1 - \frac{4}{3} \frac{\tilde{\omega}}{\gamma_a}\left(\frac{gR_a}{b}\right)^2 \xi' + \mathcal{O}(\xi')^2. \label{eq:small2}
\ee

\noi
so that the slope of the line is $-\frac{4}{3} \frac{\tilde{\omega}}{\gamma_a}\left(\frac{gR_a}{b}\right)^2$.
For the case $g = 0.7$, the slope is thus $-0.049 \tilde{\omega}$ while the reported
value\cite{rk2021} was -0.073. Thus, $\tilde{\omega} \approx 1.49$.

While Eq.~(\ref{eq:curved0}) together with Eqns.~(\ref{eq:thet}-\ref{eq:jp}) constitute a major step in
dealing with axially symmetric curved emitters, it is nevertheless an ad-hoc extension of the planar model.
The free parameter
$\tilde{\omega}$ is a reflection of this approach. Alternately, if $\tilde{\omega}$ is set to unity,
the factor $g$ which defines the average current density must be obtained from a fit to numerical results.   
There is thus sufficient scope
and motivation to build an alternate formalism which suits both
curved emitters and planar emitters with a finite active area.

\subsection{Scope for an alternate model}

A variety of alternate approaches have also been used to study
space charge effects on field emission \cite{jensen97a,jensen97b,lin2007,
rokhlenko2010,jensen2010,rokhlenko2013,jensen2014,jensen2015}.
Some of these, based on transit time, are able to reproduce  features of the time variation
of the cathode field and the steady-state that follows, especially for planar
emitters.

The field reduction factor $\tilde{\vartheta}$, which is a measure of space charge severity,
may depend on several factors for a planar emitting patch or a curved
emitter apex. In a molecular dynamics simulation\cite{torfason2015},
it was found that keeping other features of
a parallel plate diode invariant and varying only the size of the planar emitting patch,
the value of $\tilde{\vartheta}$ decreases as the size of the patch is increased. Thus, two emitters
with emitting areas $\calA_1$ and $\calA_2$ ($ > \calA_1$)
but having the same $E_L$ and transit time $T_{\tr}$, will be affected by space
charge differently with the
smaller one being relatively less affected \cite{torfason2015}.
The severity of space charge may also depend on the transit time $T_{\tr}$, the vacuum field $E_L$
and several other factors. 
For an axially symmetric curved emitters,
there are very few studies \cite{torfason2016,rk2021}. The dependence of $\tilde{\vartheta}$
on the transit time $T_{\tr}$ and the vacuum field $E_L$ are expected to persist for curved emitters.

The present communication deals with a semi-analytical model, one that
naturally accommodates curved emitters and can provide the time variation
of the apex field and emitted current and estimates of $E_P$ and $J_P$ without much
computational effort. It aims to predict some of the results reported earlier.
While it is not intended as a substitute for PIC methods, it can be used
as a design tool to scan the parameter space since a full
3-dimensional PIC modelling may be extremely time consuming and expensive.

We shall first outline the model in section \ref{sec:model} and
state the various approximations that may be used to improve the
prediction. This will be followed by a comparison with a published
MD simulation for planar geometry
and our own $\text{PIC}$ simulations using PASUPAT\cite{db_scl,db_multiscale,rk2021}.

\section{The semi-analytical model}
\label{sec:model}

\begin{figure}[hbt]
  \begin{center}
    
    \vskip -0.1cm
\hspace*{-.5cm}\includegraphics[scale=0.825,angle=0]{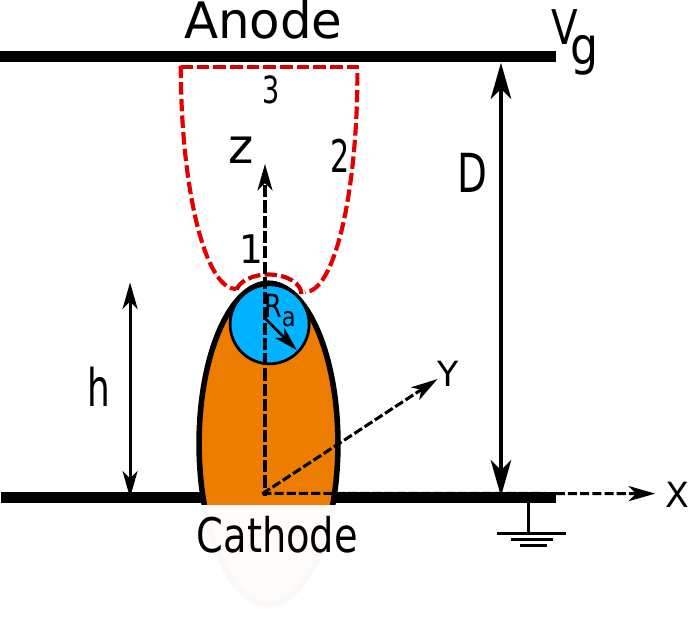}
\vskip -0.6cm
\caption{Schematic of an axially symmetric curved emitter of
  apex radius of curvature $R_a$
  and height $h$ in a parallel plate diode configuration.
  }
\label{fig:model}
\end{center}
\end{figure}

Consider an axially symmetric curved emitter of height $h$ and
apex radius of curvature $R_a$
as shown in Fig.~\ref{fig:model}. It is mounted on the grounded cathode
in a parallel plate diode configuration with the plate separation $D$,
and the anode at a potential $V_g > 0$. The macroscopic (applied) field is thus
$E_0 = V_g/D$ while the local field at the apex of the curved emitter
is $E_a(t=0) = \gamma_a E_0$ prior to the emission of electrons.
Emission can occur from any point on the curved surface (defined by $z = z(\rho)$)
depending on the strength of the local field $E_l$ at a point ($\rho,z$)
on the surface.

After emission begins ($t > 0$), the space between the plates has an amount
of charge $Q(t)$ while the local field at the apex is $E_a(t)$. Since the
field emission current density\cite{MG,forbes2006} $J(t)$ depends on the local
field $E_l(\rho)$, the emitted current can vary with time. For $t < T_{\tr}$, where
$T_{\tr} = D/(\sqrt{2eV_g/m}/2)$ is the ballistic transit time,
$Q(t) = Q_{\inj}(t)$ where

\bea
  Q_{\inj} (t) & = & \int_0^t dt' I_{\inj}(t') \label{eq:charge_in} \\
 I_{\inj} & = & \int_S J(E_l(\rho,t'))~2\pi\rho \sqrt{1 + (dz/d\rho)^2} d\rho \label{eq:J}
\eea

\noi
where $S$ denotes the surface of the curved emitter $z = z(\rho)$ and the field
emission current density

\be
J =  \frac{1}{\mathlarger{t}_F^2(\rho)} \frac{A_\fn}{\phi} E_l(\rho)^2 \exp(-B_\fn v_{{\small F}}(\rho) \phi^{3/2}/E_(\rho)). \label{eq:MG}
\ee

\noi
with

\bea
v_{\small F} & \simeq & 1 - f + \frac{1}{6} f \ln f \\
t_{\small F} & \simeq & 1 + \frac{f}{9} - \frac{1}{18} f \ln f \\
f & \equiv & c_S^2 \frac{E_l(\rho)}{\phi^2} \label{eq:last}
\eea

\noi
where  $A_\fn~\simeq~1.541434~{\rm \mu A~eV~V}^{-2}$,
$B_\fn~\simeq 6.830890~{\rm eV}^{-3/2}~{\rm V~nm}^{-1}$ are the conventional Fowler-Nordheim constants,
$c_S$ is the Schottky constant with $c_S^2 = 1.439965~ \text{eV}^2 \text{V}^{-1} \text{nm}$ and
$\phi$ is the work function of the material under consideration in eV.
If the apex radius of curvature is smaller than $100$nm, it is advisable to use the
curvature-corrected form of the field emission current density\cite{DB_RR,db_rr_2021}.

Note that $E_l(\rho)$ depends on time when emission is space charge affected and
so is the current density $J$.
In general, it suffices
to take the limits of $\rho$ integration as $[0,R_a]$. For
$t > T_{\tr}$, electrons leave the diode region so that $Q(t)$ can decrease
leading to an increase in $E_a(t)$. The loss of electrons can be approximately
modelled as

\be
Q_{out}(t) = \Theta(t - T_{\tr}) \int_0^t~dt'~ I_{\inj}(t' - T_{\tr}) \label{eq:charge_out}
\ee

\noi
so that the electrons emitted at $t - T_{\tr}$ arrive at the anode at time $t$.
In the above $\Theta(t - T_{\tr})$ is the Heaviside step function which is zero
for $t < T_{\tr}$ and equals 1 for $t > T_{\tr}$.
Thus, at any time $t$, the amount of charge in the diode is

\be
Q(t) = Q_{\inj}(t) - Q_{out}(t).  \label{eq:charge}
\ee

\noi
An oscillatory behaviour in $Q(t)$ is thus to be expected, especially if the field
at the emitter apex 
falls substantially from its vacuum value $E_L = E_a(t=0)$ leading to periods
of low current injection.

Note that the local field $E_l(\rho,t)$ at any point ($\rho,z(\rho)$) close to the apex, is related to $E_a(t)$
through the generalized cosine law\cite{cosine1,cosine2}

\be
E_l(t) = E_a(t) \cos\tilde{\theta} = E_a \frac{z/h}{\sqrt{(z/h)^2 + (\rho/R_a)^2}}
\ee

\noi
which has been found to hold for emitters with parabolic endcaps
obeying $z \approx h - \rho^2/(2R_a)$. This local variation holds in the absence of
space charge as shown analytically (using the nonlinear line charge model)
as well as numerically. It has recently been demonstrated that the cosine variation
holds for moderate space charge intensity with reasonable accuracy\cite{rk2021}.

A relation between $E_a(t)$ and $Q(t)$ can be obtained by applying Gauss's law twice,
first at $t = 0$ and then at some arbitrary time $t$. Consider a Gaussian surface
denoted by the dashed closed curve marked 1,2,3 in Fig.~\ref{fig:model}.
At time $t = 0$, the field close to the anode (surface 3) is $-E_A \hat{z} = -E_0\hat{z}$
while close to the emitter-apex (on surface 1), the field is $- \gamma_a E_0 \hat{z}$.
Elsewhere on surface 1, the magnitude of the electric field is assumed to vary
accordingly to the generalized cosine law so that the total flux passing through
surface 1 may be evaluated.

If we further choose surface 2 such that the field lines lie on it (rather than
intersecting it any point), the flux through surface 2 is zero. Also, since there is no
charge present at time $t = 0$,

\be
\int_{S_1} \vec{E}.\vec{dS} +  \int_{S_2} \vec{E}.\vec{dS} + \int_{S_3} \vec{E}.\vec{dS} = 0 
\ee

\noi
which implies $E_A \calA = \int_{S_1} \vec{E}.\vec{dS}$. The flux through the parabolic surface 1 may
be evaluated as follows:

\bea
&\;& \int_{S_1} \vec{E}.\vec{dS}~\simeq~\cf \int_0^1 d\ttr \sqrt{1 + \ttr^2} \frac{1 - \ttr^2 R_a/(2h)}{\sqrt{1 - \ttr^2 R_a/h + \ttr^2}} \nonumber \\
& \simeq & \cf \int_0^1 d\ttr~ \ttr \left( 1 + \frac{\ttr^2}{1+\ttr^2}\frac{R_a}{2h} + \dots \right)\left(1 - \frac{\ttr^2}{2}\frac{R_a}{h}\right) \nonumber \\
& = & \frac{\cf}{2} \left[1 - \frac{R_a}{h} \int_0^1 d\ttr~\frac{\ttr^4}{1 + \ttr^2} \right] \nonumber \\
& = & \gamma_a E_0 \pi R_a^2 \left[ 1 - \frac{R_a}{h} \left(\frac{\pi}{4} - \frac{2}{3} \right) \right]
\eea

\noi
where $\ttr = \rho/R_a$ and $\cf = \gamma_a E_0 2\pi R_a^2$. For a sharp emitter
where $R_a/h << 1$, the flux is
$\gamma_a E_0 \pi R_a^2$. In general, $E_A \calA =\gamma_a E_0 \pi R_a^2 [1 - (R_a/h)(\pi/4 - 2/3)]$
and since $E_A = E_0$ at $t = 0$,

\be
\calA(t=0) = \calA(0) =  \gamma_a \pi R_a^2  [1 - (R_a/h)(\pi/4 - 2/3)]. \label{eq:area_anode}
\ee

\noi
Thus for a curved emitter, the flux tube expands on reaching the anode.

We can now consider the case when the Gaussian surface encloses an amount of charge $Q$
and apply Gauss's law again. If we assume the generalized cosine law to be approximately
valid in the presence of space charge,

\be
\int_{S_1} \vec{E}.\vec{dS} = E_a(t) \pi R_a^2 \left[ 1 - \frac{R_a}{h} \left(\frac{\pi}{4} - \frac{2}{3} \right) \right]
\ee

\noi
where $\gamma_a E_0 = E_a(0)$ has been replaced by $E_a(t)$. Surface 2 can again be chosen such that
field lines do not cross it and hence $\int_{S_2} \vec{E}.\vec{dS} = 0$. Finally, we recognize the
fact that surface 3 (infinitesimally close to the anode) may be larger on account of space charge
and the field $E_A$ may be significantly larger in magnitude than $E_0$. We shall denote the
area and anode field by $\calA(t)$ and $E_A(t)$ respectively and their asymptotic steady-state
values by $\calA(\infty)$ and $E_A(\infty)$.

The unknown quantities are thus (i) $E_a(t)$ which is the field at the apex of the curved emitter
(ii) $E_A(t)$ which is the field at the flat anode (iii) $\calA(t)$ which is the
area through which is the anodic-flux reaches the curved emitter tip from the apex to $\rho = R_a$.
While $E_A(t)$ and $\calA(t)$ are required to refine the calculation of $E_a(t)$, we can
assume as a first approximation that they assume their vacuum values i.e. $E_A(t) = E_A(0) = E_0$ while
$\calA(t) = \calA(0)$.

With this first approximation, an application of Gauss's law leads us to the equation

\be
E_a(t) \pi R_a^2 \left[ 1 - \frac{R_a}{h} \left(\frac{\pi}{4} - \frac{2}{3} \right) \right]
= E_A(0) \calA(0) - Q(t)
\ee

\noi
where $Q(t)$ is given by Eq.~(\ref{eq:charge}). Substituting $E_A(0) = E_0$ and $\calA(0)$ with the
expression in Eq.~(\ref{eq:area_anode}), we have

\be
E_a(t) = E_a(0) - \frac{1}{\calA_C} Q(t)  \label{eq:setup1}
\ee

\noi
where $E_a(0) = \gamma_a E_0$ and the effective area at the curved emitter

\be
\calA_C =  \pi R_a^2 \left[ 1 - \frac{R_a}{h} \left(\frac{\pi}{4} - \frac{2}{3} \right) \right].
\ee

\noi
Since $Q(t)$ depends on the field $E_a(t)$, Eq.~(\ref{eq:setup1}) can be differentiated to
yield

\be
\frac{dE_a}{dt} = -\frac{1}{\calA_C} \left[ I_{\inj}(t) - \Theta(t - T_{\tr}) I_{\inj}(t - T_{\tr}) \right] \label{eq:setup2}
\ee

\noi
where $I_{\inj}(t)$ is given by Eq.~(\ref{eq:J}).
Eq.~(\ref{eq:setup2}) can be solved with the initial
condition $E_a(0) = \gamma_a E_0$.

It is clear that the injection of charges into the diode and their subsequent loss at
the anode leads to an oscillatory evolution of the apex field $E_a(t)$ if
emission falls substantially from the initial values.

\subsection{The anode field approximation}

In order to improve the predictive power of the model, it is important to approximate
the anode field $E_A(t)$. At $t = 0$, when the field at the emitter is $\gamma_a E_0$,
the anode field is $ E_0$. We are also aware from planar space charge
limited flows that when the field at the cathode is zero (the limiting case),
the anode field assumes the value $4E_0/3$. A simple linear interpolation between
these two points $(\gamma_a E_0,E_0)$ and $(0,4E_0/3)$ on the
$(E_a,E_A)$ plane, gives us the relation

\be
E_A(t) = -\frac{1}{3\gamma_a} E_a(t) + \frac{4}{3}E_0.  \label{eq:anode}
\ee

\noi
Since $E_a$ varies with time, the dependence on time is explicitly shown.
Note that Eq.~(\ref{eq:anode}) reproduces the two limits since $E_a(0) = \gamma_a E_0$
and the anode field assumes the value $4E_0/3$ when $E_a(t) = 0$.

If we consider the anode area $\calA(t)$ to be invariant in time and assume its vacuum value,
an application of Gauss's law with only the anode-field correction leads us to the equation

\be
\begin{split}
E_a(t) \pi R_a^2 & \left[ 1  - \frac{R_a}{h} \left(\frac{\pi}{4} - \frac{2}{3} \right) \right]
= \\
& \calA(0) \left[ \frac{4}{3}E_0 - \frac{1}{3\gamma_a} E_a(t) \right]   - Q(t)
\end{split}
\ee

\noi
so that on rearranging terms

\be
\frac{4}{3} E_a(t) \calA_C = \frac{4}{3} E_a(0) \calA_C - Q(t)
\ee

\noi
and differentiating, we have

\be
\frac{dE_a}{dt} = -\frac{3}{4\calA_C} \left[ I_{\inj}(t) - \Theta(t - T_{\tr}) I_{\inj}(t - T_{\tr}) \right]. \label{eq:setup3}
\ee

\noi
Thus, the anode-field correction introduces a factor $3/4$ in the rate at which the apex
field changes.

\subsection{The anode area approximation}

At the next level, we can also introduce an anode area approximation using the parallel-plate diode as
a guide. Before emission starts, the flux tube maps equal area at the cathode and anode in a planar
diode while in case of a curved emitter, the flux tube maps an area of the cathode that is $\gamma_a$ times more
at the anode (see Eq.~\ref{eq:area_anode}). Thus, when the field at the emitter apex is $E_a(0)$, the area at anode is $\gamma_a \calA_C$.

\begin{figure}[hbt]
  \begin{center}
    
    \vskip -0.1cm
\hspace*{-.5cm}\includegraphics[scale=0.825,angle=0]{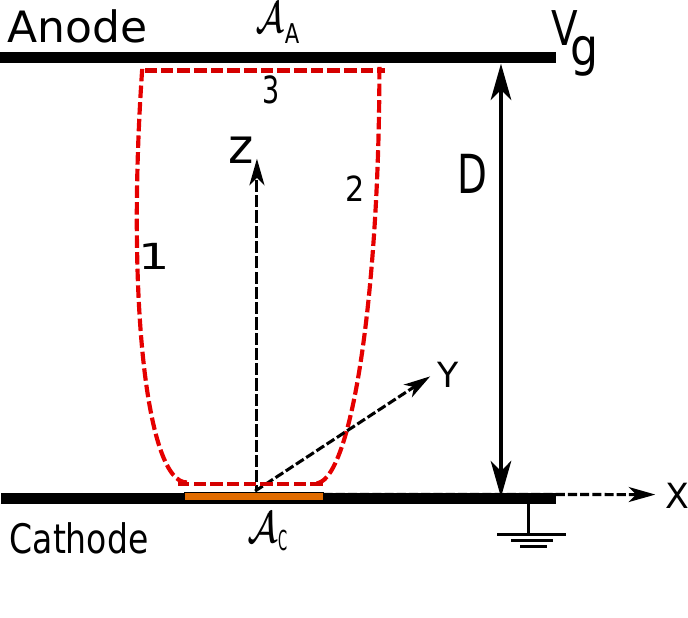}
\vskip -0.6cm
\caption{Schematic of a finite-sized planar emitting patch of area $\calA_C$
  in a parallel plate diode configuration along
  with a Gaussian surface (dashed curve) mapping $\calA_C$ to $\calA_A$ at the anode.
  }
\label{fig:model2}
\end{center}
\end{figure}

The planar Child-Langmuir law can again be used to obtain a second point. Numerical and analytical results show
that if the emitting patch is finite, the Child-Langmuir current density can be expressed as
$J_{\CL} (1 + \alpha D/W)$ where $W$ represents the size of the emitting patch and $\alpha$ is a constant
that depends on the shape of the patch. Thus, for a circular patch, $\alpha = 1/4$ while $W$ is its
radius.
The relevance of the factor $(1 + \alpha D/W)$ in the planar case with finite emission area
(see Fig.~\ref{fig:model2}) can be
understood by considering the Gaussian surface as a 
the flux tube extending from the cathode having an area $\calA_C$, to the anode having an area $\calA_A$.
On applying Gauss's law and assuming that the cathode field is zero, we have
$-E_A \calA_A = -Q/\epsilon_0$ where $Q = I_{\CL} T_{\tr} = J_{\CL}( 1 + \alpha D/W) \calA_C T_{\tr}$.
Assuming $E_A = 4E_0/3$  and $T_{\tr} = D/v_{av}$
where the average speed $v_{av} = v_{max}/3 = \sqrt{2e V_g/m}/3$ as in the infinite parallel plate case, it follows that
$\calA_A = ( 1 + \alpha D/W) \calA_C$. In other words, the area of the emitting patch at the
cathode is mapped to a patch at the anode that is larger by a factor $( 1 + \alpha D/W)$.
When applied to a curved emitter having an effective emitting area $\calA_C$, the
area at the anode is $\gamma_a \calA_C (1 + \alpha D/W)$ when the field at the apex is zero.

A linear interpolation between these two points gives

\be
\calA_A(t) = \gamma_a \calA_C \left[ -\frac{\alpha D/W}{\gamma_a E_0} E_a(t) + (1 + \frac{\alpha D}{W})\right] . \label{eq:area_approx}
\ee

\noi
Note that Eq.~(\ref{eq:area_approx}) states that at $t=0$ when $E_a(0) = \gamma_a E_0$,
$\calA_A = \gamma_a \calA_C$ while when $E_a = 0$, $\calA_A = (1 + \alpha D/W) \gamma_a \calA_C$.
For an axially symmetric curved emitter, $W$ may be taken to be $R_a$ and $\alpha = 1/4$.

An application of Gauss's law with both the anode field and area corrections leads to the
equation

\be
\begin{split}
E_a(t) \calA_C &
= \left[ \frac{4}{3}E_0 - \frac{1}{3\gamma_a} E_a(t) \right] \times \gamma_a \calA_C \times \\
& \left[ -\frac{\alpha D/W}{\gamma_a E_0} E_a(t) + \left(1 + \frac{\alpha D}{W}\right)\right]   - Q(t)
\end{split}
\ee

\noi
which can be rearranged and differentiated to yield

\be
\frac{dE_a}{dt} \left[\frac{4}{3} + \alpha \frac{D}{W} - \alpha \frac{2}{3}\frac{D}{W} \frac{E_a(t)}{E_a(0)} \right] = -\frac{I(t)}{\calA_C}  \label{eq:setup4}
\ee

\noi
where

\be
I(t) = \left[ I_{\inj}(t) - \Theta(t - T_{\tr}) I_{\inj}(t - T_{\tr}) \right]. \label{eq:netcurrent}
\ee

Eq.~(\ref{eq:setup2}), Eq.~(\ref{eq:setup3}) and (\ref{eq:setup4}) along with Eq.~(\ref{eq:netcurrent})
provide successively better approximations for determining the steady state electric field
$E_P = E_a(\infty)$ at the emitter apex.
Note that the value of $\alpha$ is not known with any accuracy and may need to be determined by
fitting Particle-in-Cell or Molecular Dynamics data, especially for curved emitters.
Nevertheless, it is expected to provide a fast approximate determination of the space
charge affected field emission current.

The model presented here is also applicable to a cluster of emitters or a large area field emitter (LAFE).
The hybrid model proposed recently\cite{db_rudra1,rudra_db,db_rudra2,anodeprox,db_hybrid} can be
used to determine the apex field enhancement factor of individual emitters in the LAFE before emission starts,
and so long as  two emitters are not too close to each other for mutual space charge
effects to kick in, the formalism presented here can be used for each emitter.

\section{Numerical Results}

We shall use Eq.~(\ref{eq:setup4}) along with Eq.~(\ref{eq:netcurrent}), Eq.~(\ref{eq:J}) and
Eq.~(\ref{eq:MG}) to test the
usefulness of the theoretical model proposed in Section \ref{sec:model}. While, we expect the gross
features of the time evolution to be visible, finer details are not expected since they are beyond the
scope of the approximations used. It would be interesting to see if the steady state field $E_P$ is
determined consistently with reasonable accuracy.

\subsection{Comparison with planar result}

A first test of the model is the extensive Molecular Dynamics (MD) data reported in Ref.~[\onlinecite{torfason2015}]
for a square emitting patch of side length $L$ in a planar diode geometry. The gap between the anode
and cathode plates is $1000$nm, the potential difference $V_g = 2$kV while $L$ varies from
$50$nm to $2500$nm. The emission current density follows Eq.~(\ref{eq:MG})
with a work function $\phi = 2$eV. Since the emitting area is a flat square, $E_a(t)$ refers to
the field on the emitter surface. The vacuum field $E_L$ is thus $2$V/nm.
When emission starts, it is assumed that the field on the emitter is
independent of the location on the patch resulting in uniform emission (in reality, the wings have larger
current density especially when $L$ is small).
Thus $I_{\inj} = L^2 J$ where $J$ is computed using Eq.~(\ref{eq:MG}).

\begin{figure}[hbt]
  \begin{center}
    \vskip -0.1cm
\hspace*{-.5cm}\includegraphics[scale=0.325,angle=0]{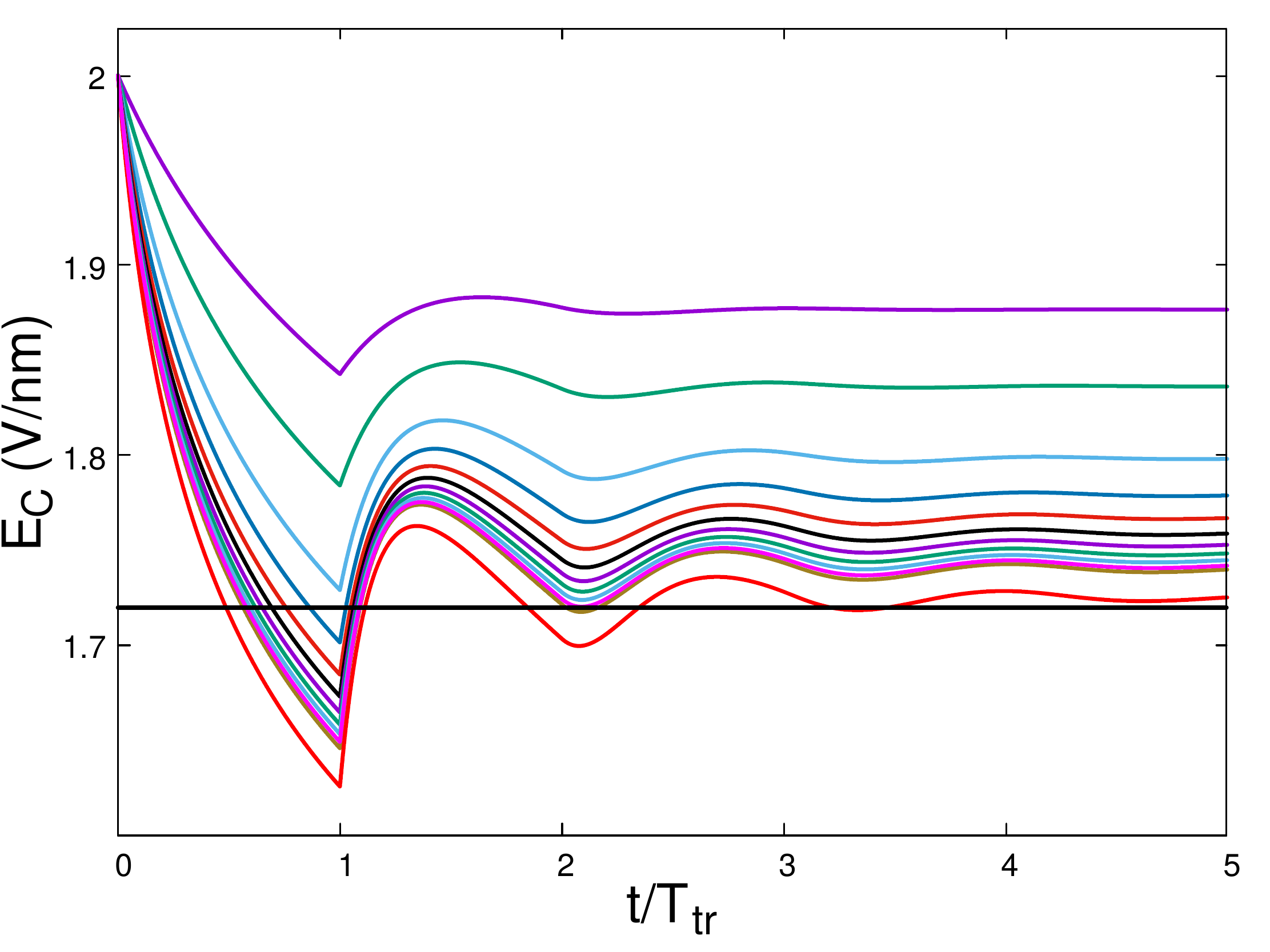}
\caption{Time evolution for the cathode field in units of the
  transit time $T_{\tr}$, for different emitting-patch size $L$ .
  The bottom most oscillatory curve is for $L = 2500$nm while the
  topmost is for $L = 50$nm. The value of $\alpha = 1$. The horizontal
  line marks the PIC result $E_C = 1.72$V/nm.
  }
\label{fig:Ea_planar}
\end{center}
\end{figure}

When the patch length is large ($L >> D$), the results are expected to mimic the 1-D PIC
result reported in Ref.~[\onlinecite{feng2006}] where the space-charge affected cathode field
is around $1.72$V/nm. The MD simulation\cite{torfason2015} indeed approaches this value when $L = 2500$nm.
For the model presented here, the anode-area correction is expected to
be small for $L = 2500$nm and the results should not be very sensitive on the value of $\alpha$. This is
indeed found to true with very little variation as $\alpha$ varies from nearly 0 to 1. 
The smaller $L$ values are however poorly reproduced for standard values of $\alpha$
in the range $[1/4,\sqrt{2}/\pi]$ with $\sqrt{2}/\pi$ being the value appropriate for a square
of side-length $L$.
Fig.~(\ref{fig:Ea_planar}) shows a plot of the cathode field variation for $\alpha = 1$ for values of
$L = 2500,1000,900,\ldots,100,50$nm (bottom to top curves). While these do not match perfectly with
the MD simulation results (see Fig.~4 of Ref.~[\onlinecite{torfason2015}]) especially for $L < 500$nm,
the trend is very nearly the same and the error is reasonably small.

\subsection{Curved emitters}

Planar field emitters require high macroscopic fields in the $E_0 = 3-10$ V/nm range due to the lack of
field enhancement ($\gamma_a = 1$). In practice, field emission occurs from specially designed curved emitters
or nano-protrusions on a smooth surface since they require a much smaller macroscopic
field. These are thus the natural geometries that need to be investigated for the effect of space
charge on field emission.

We shall consider here an axially symmetric curved
emitter for which the projected emission area can be considered to be circular of radius $R_a$ and the free parameter
$\alpha$ can be chosen to be equal to $1/4$. The hemi-ellipsoid is an example
of such a geometry. While, the Laplacian problem for the hemi-ellipsoid placed
on a conducting plane with the anode far away can be solved analytically \cite{smythe,kosmahl},
the presence of space charge makes the problem non-trivial and requires numerical investigation.
In order to test our model, we shall consider various magnifications of the basic hemi-ellipsoidal
emitter having a height $h = 2.515\mu$m and base radius $b = 1.5\mu$m.
The apex radius of curvature $R_a = b^2/h \simeq 0.8946\mu$m.
The spacing between the anode and cathode plates is $D = 10\mu$m.
The work function is considered to be $\phi = 4.5$eV.

\begin{figure}[hbt]
  \begin{center}
    \vskip -0.85cm
\hspace*{-.5cm}\includegraphics[scale=0.34,angle=0]{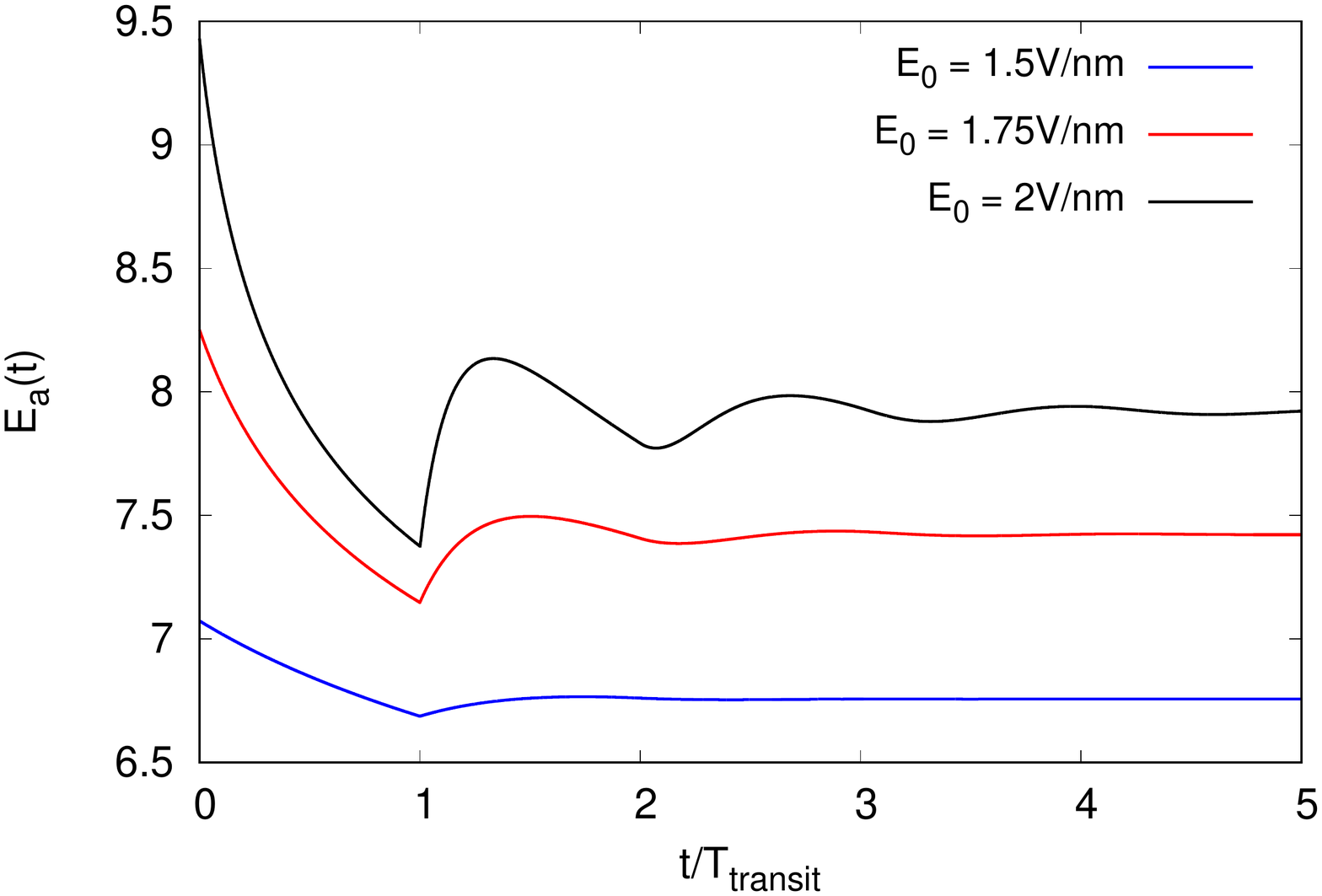}
\vskip -0.4in
\caption{The time variation of the apex field determined using the model for three values of $E_0$. Note that
  the oscillation gets pronounced as the field increases and is negligible at $E_0 = 1.5$V/nm.
The spacing between the anode and cathode plates is $D = 10\mu$m. The parameter $\alpha = 1/4$.
}
\vskip -0.15in
\label{fig:Ea_D10_model}
\end{center}
\end{figure}

The predictions of the model for $V_g = 15, 17.5$ and $20$kV corresponding to $E_0 = 1.5, 1.75$ and $2$V/nm
are shown in Fig.~\ref{fig:Ea_D10_model}. While space charge affects the apex field $E_a$ at all three
values of the macroscopic field, it is strong enough to causes oscillations at $E_0 = 1.75$ and $2$V/nm.
The nature of oscillations is similar to the planar case with the period linked to the transit time
as observed in planar molecular dynamics simulations\cite{torfason2015} and theoretical models\cite{jensen2015}.
Planar PIC simulations also display oscillations\cite{feng2006} but these have not been observed yet
in 3-D simulations using curved emitters \cite{edelen2020} which are considerably more resource intensive. 

\begin{figure}[thb]
  \begin{center}
    \vskip -0.65cm
\hspace*{-.5cm}\includegraphics[scale=0.34,angle=0]{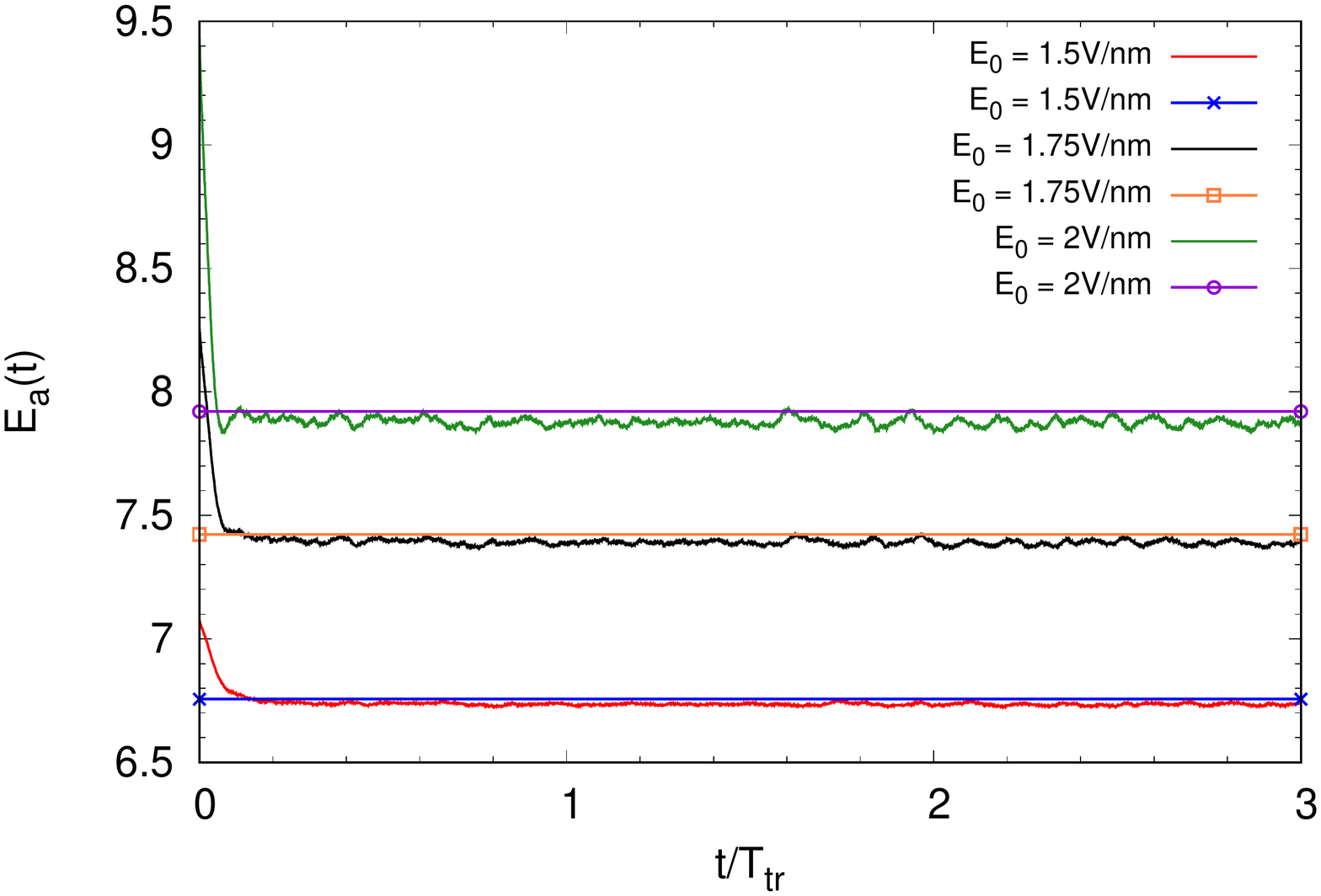}
\vskip -1.0cm
\caption{Time variation of the apex field for $E_0 = 1.5, 1.75$ and $2.0$V/nm. The straight line marks the
  steady-state value obtained using the model with $\alpha = 1/4$.
  }
\label{fig:Ea_D10}
\end{center}
\end{figure}

\begin{figure}[hbt]
  \begin{center}
      \vskip -1.5cm

\hspace*{-.5cm}\includegraphics[scale=0.34,angle=0]{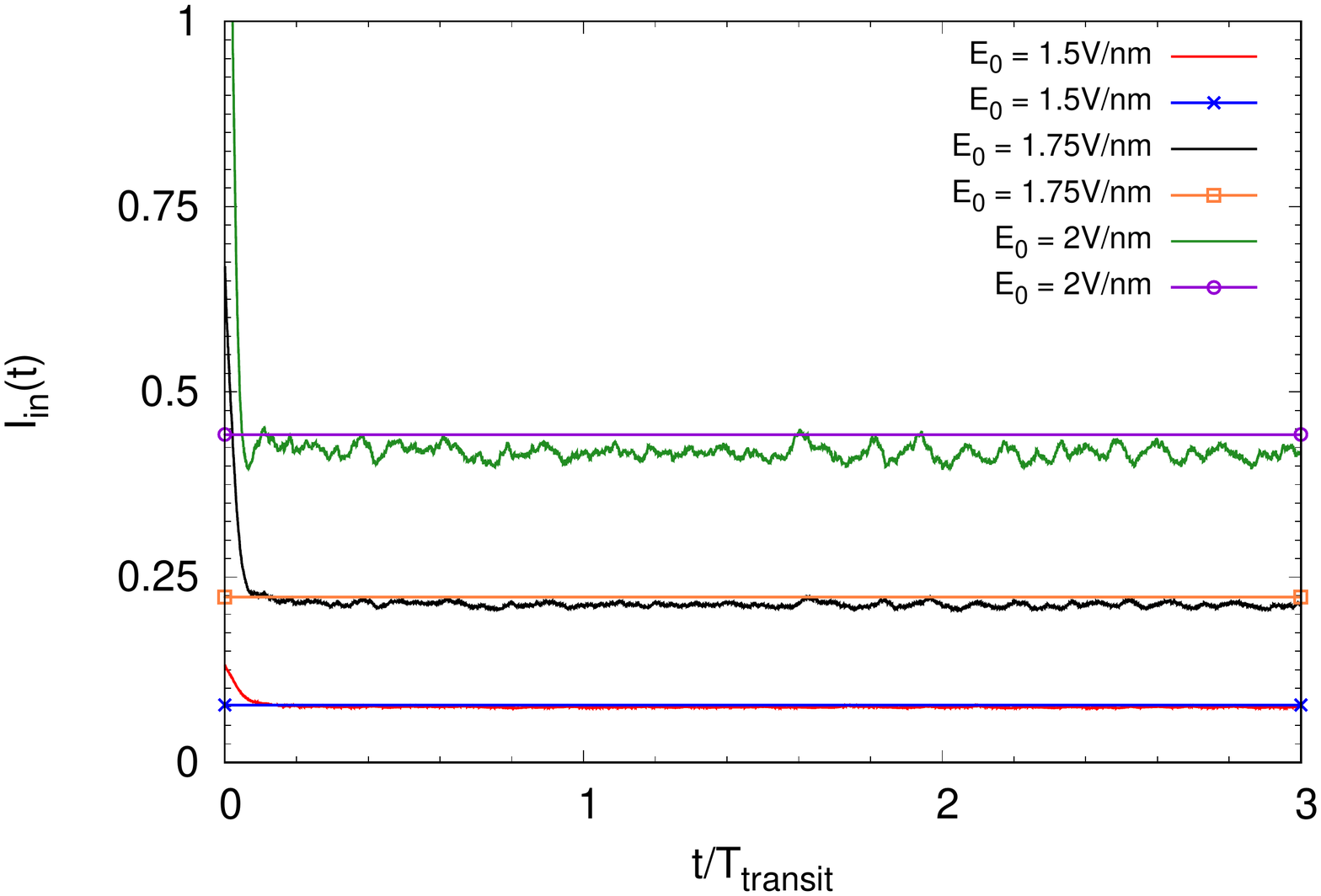}
\vskip -1.0cm
\caption{Time variation of the injected current for $E_0 = 1.5, 1.75$ and $2.0$V/nm. The straight line marks the
  steady-state value  obtained using the model with $\alpha = 1/4$.
}
\vskip -0.1in
\label{fig:I_D10}
\end{center}
\end{figure}

We shall henceforth compare the model predictions with PIC simulations performed using PASUPAT \cite{db_scl,db_multiscale,rk2021}
with a field emission module based on the cosine law \cite{db_parabolic,db_multiscale,rk2021}.
Our focus will be on the steady state values of the apex field and emitted current and these will be
compared with the predictions of the model for various geometric diode parameters.
In the PIC simulation using PASUPAT, the hemiellipsoid and cathode
plate are considered to be grounded  perfect electrical conductors (PECs) while the anode is a PEC at a voltage
$V_g$. The centre of the hemiellipsoid is at $(X,Y) = (0,0)$ while the transverse boundaries are located at
$X,Y = \pm 5 \mu$m and have Neumann boundary condition imposed on them. Since the height of the emitter
is small compared to the extent of the boundary in the $X$ and $Y$ directions, it is close to being
an isolated emitter. The value of $\gamma_a$ as calculated using PASUPAT is 4.715 which coincides with
the value evaluated using COMSOL. The apex field enhancement effect for an isolated emitter is
around $4.8$. Thus, there is a mild shielding effect due to the computational boundary not being very
far away.

For the diode described above with $D = 10\mu$m, the PIC results for $E_0 = 1.5, 1.75$ and $2$V/nm
are shown in Fig. \ref{fig:Ea_D10}. The straight line in each case marks the prediction of the model presented
in section \ref{sec:model}
for the steady-state apex field. The corresponding result for the emission current is
shown in Fig. \ref{fig:I_D10}. The agreement is reasonably good especially at the lower values of
the macroscopic field where the agreement with the cosine law is good\cite{rk2021}. Note that
in order to keep the fluctuations small, the number of time steps per transit time ($\approx 3000)$ has been
kept identical for all the simulations.

\begin{figure}[hbt]
  \begin{center}
      \vskip -0.75cm
\hspace*{-.5cm}\includegraphics[scale=0.34,angle=0]{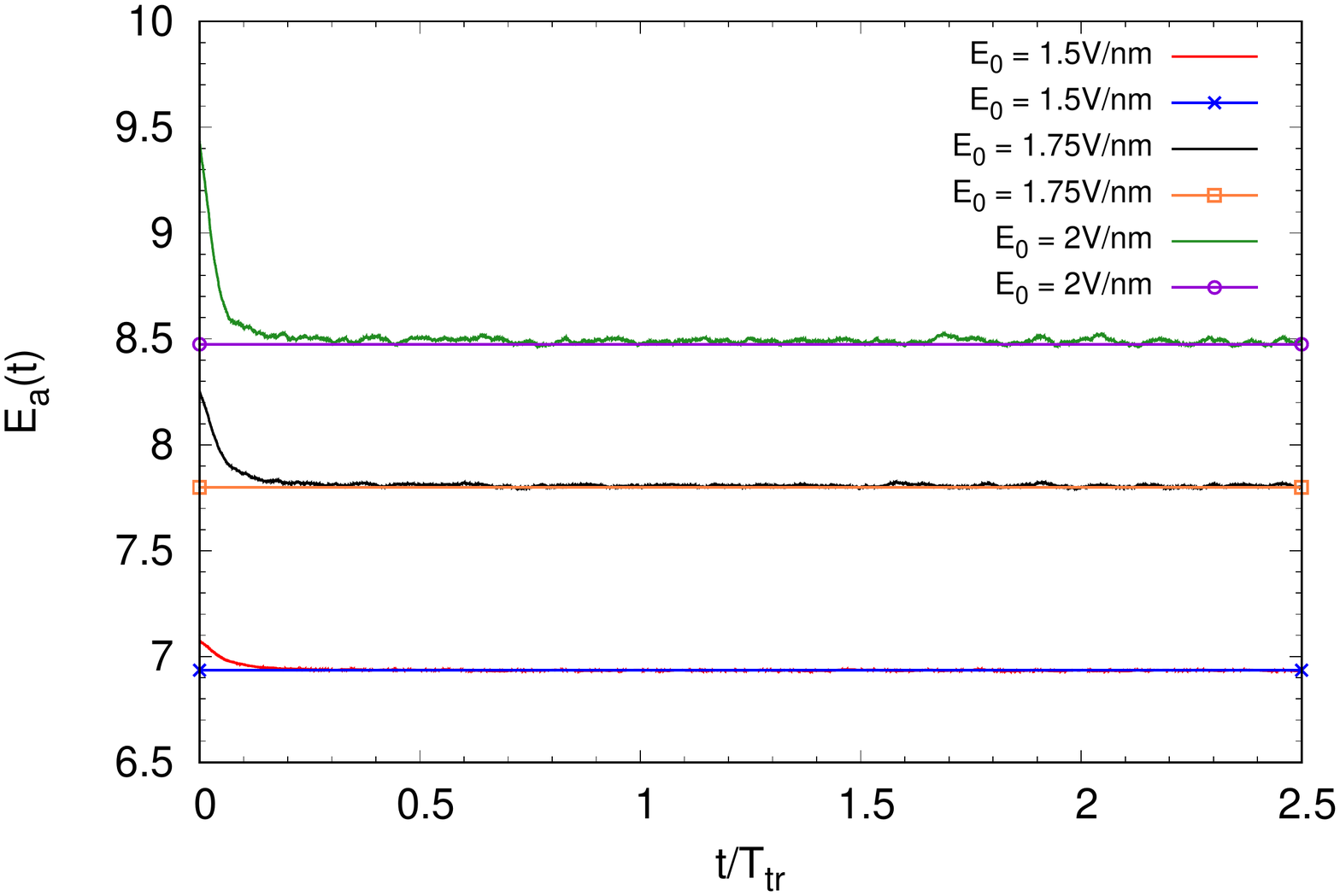}
\vskip -0.9cm
\caption{Time variation of the apex field for the diode scaled down by a factor of 10.
  Thus, $D = 1\mu$m. The straight line marks the
  steady-state value obtained using the model with $\alpha = 1/4$.
}
\vskip -0.25in
\label{fig:Ea_D1}
\end{center}
\end{figure}

\begin{figure}[hbt]
  \begin{center}
      \vskip -0.75cm
\hspace*{-.5cm}\includegraphics[scale=0.335,angle=0]{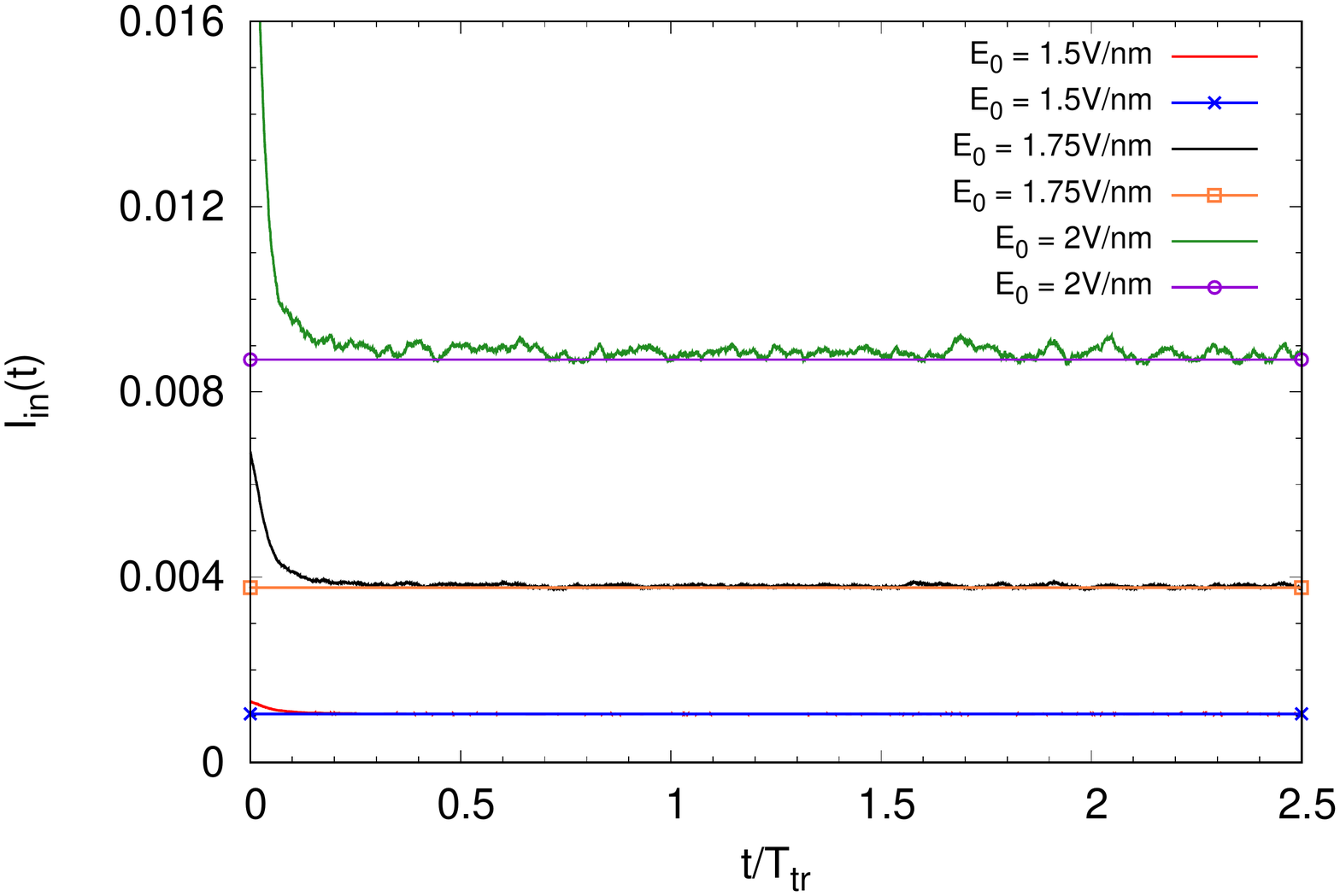}
\vskip -0.6cm
\caption{The corresponding injected current variation. with time The straight line marks the
  steady-state value obtained using the model with $\alpha = 1/4$.
  }
\label{fig:I_D1}
\end{center}
\end{figure}

We shall next consider the effect of transit time by scaling the size of the diode while
keeping the field at the apex $E_a(0)$ as invariant. We shall first consider a scaling down
of all geometric quantities by a factor of 10. Thus, $h = 2.515\times 10^{-1}\mu$m, $b = 1.5\times 10^{-1}\mu$m
$D = 10 \times10^{-1}\mu$m and the transverse computational boundaries are at $X,Y = \pm 5 \times 10^{-1}\mu$m.
In terms of $E_0$, the transit time $T_{\tr} \sim (D/E_0)^{1/2}$. Since the enhancement factor $\gamma_a$
is unchanged by the scaling, $E_a(0)$ remains the same if $E_0$ is maintained at the previous values.
Thus $T_{\tr} \sim \sqrt{D}$. Thus, on scaling down  the diode by a factor of 10, the
transit time decreases by a factor of $\sqrt{10}$ resulting in faster loss of charges from the diode.
The effect of space charge is thus expected to be weaker.
The PIC simulation results for the scaled down diode are shown in Fig.~\ref{fig:Ea_D1}
and Fig.~\ref{fig:I_D1}. A comparison with Fig.~\ref{fig:Ea_D10} shows
the drop in field from the vacuum values to be
much smaller indicating a larger value of $\vartheta$. Note that the agreement with the
cosine law in the presence of space charge is excellent for this scaled-down diode\cite{rk2021}.

\begin{figure}[hbt]
  \begin{center}
    \vskip -0.75cm
\hspace*{-.5cm}\includegraphics[scale=0.34,angle=0]{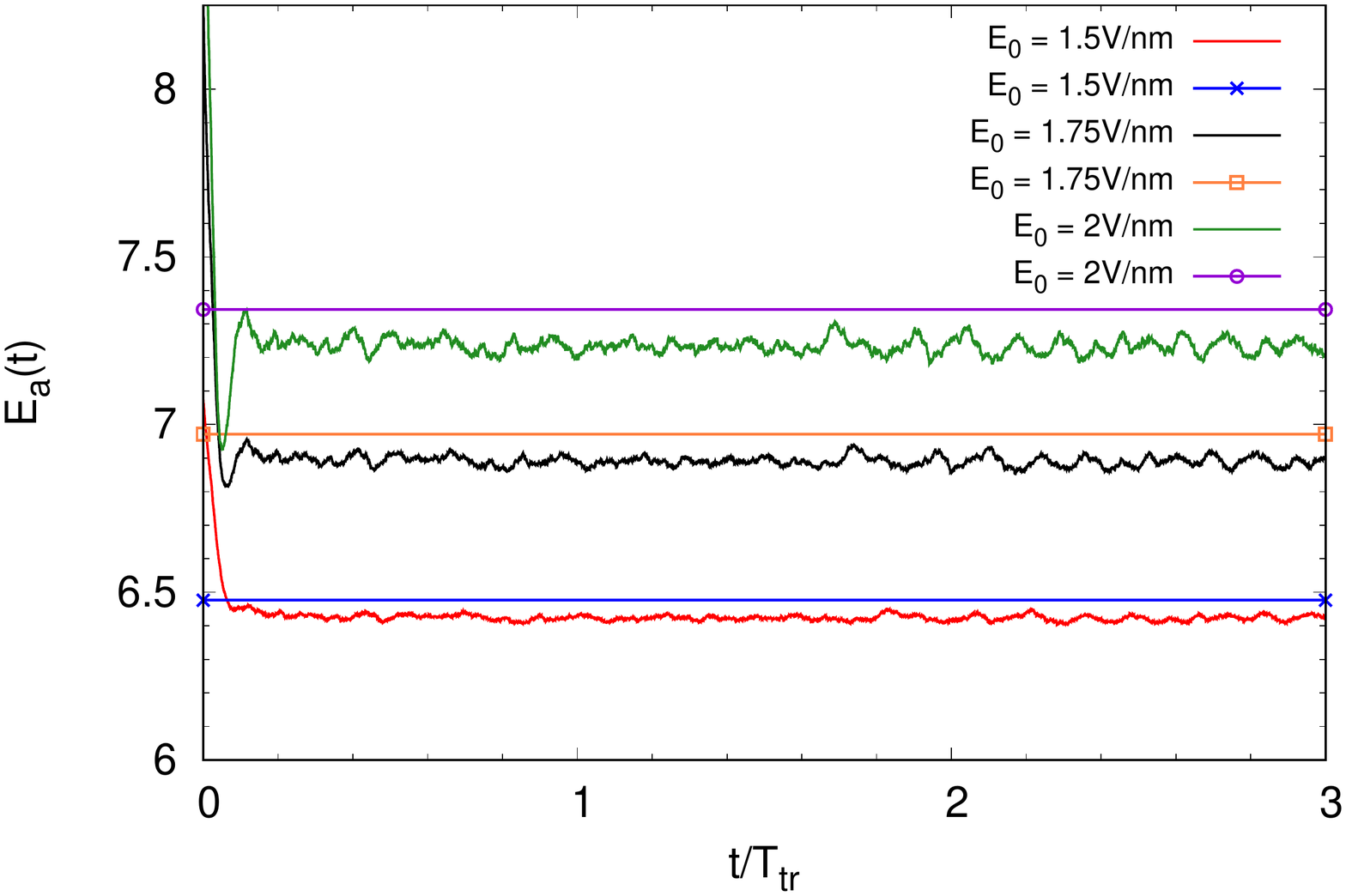}
\vskip -1.25cm
\caption{Time variation of the apex field for the scaled-up diode with $D = 100 \mu$m.
  Also shown by straight lines are the predictions for the steady-state field using the
  model.
  }
\label{fig:Ea_D100}
\end{center}
\end{figure}

\begin{figure}[hbt]
  \begin{center}
    \vskip -0.75cm
\hspace*{-.5cm}\includegraphics[scale=0.34,angle=0]{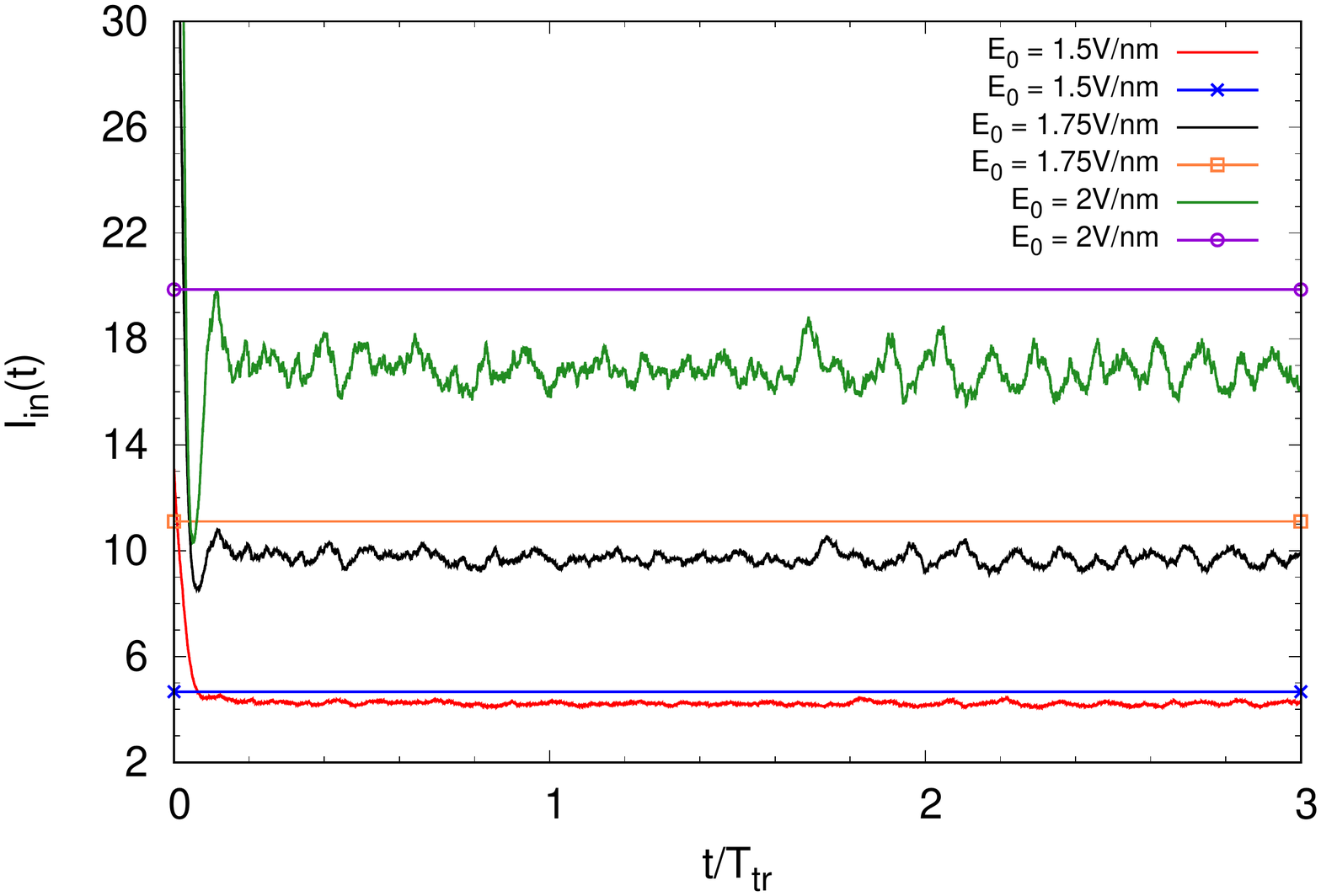}
\vskip -0.6cm
\caption{The corresponding time variation of injected current for the scaled-up diode. 
  }
\label{fig:I_D100}
  \end{center}
\end{figure}

We next consider a scaled-up diode to check for the consistency of our results.
The scaling factor is 10 so that $h = 2.515\times 10\mu$m, $b = 1.5\times 10\mu$m
$D = 10 \times10 \mu$m and the transverse computational boundaries are at
$X,Y = \pm 5 \times 10 \mu$m. The results for the time variation of the apex
field and injected current are shown in Figs.~\ref{fig:Ea_D100} and \ref{fig:I_D100}.
The effect of space charge is much stronger as compared to the unscaled
case ($D = 10\mu$m) thus establishing the importance of transit time
in determining the severity of space charge effect on field emission.
Not surprisingly, even at $E_0 = 1.5$V/nm, the agreement between the
model and the PIC result is not perfect. This coincides with the
larger deviation from the cosine law for the scaled-up diode\cite{rk2021}.

\begin{figure}[hbt]
  \begin{center}
    \vskip -0.75cm
\hspace*{-.5cm}\includegraphics[scale=0.34,angle=0]{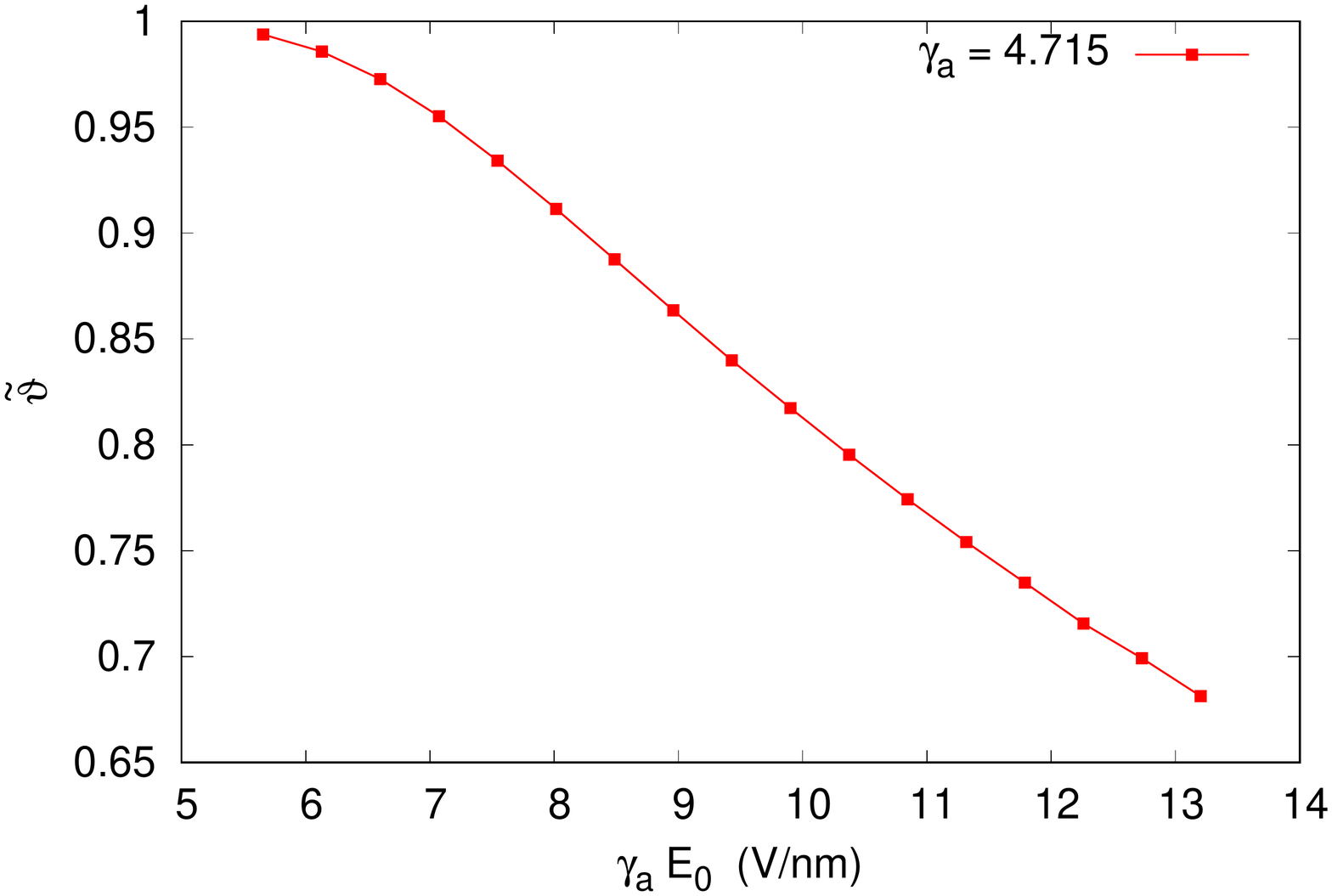}
\vskip -0.6cm
\caption{The field reduction factor $\vartheta$ obtained from the model is plotted
  against the apex vacuum field $E_a(0) = \gamma_a E_0$ for $D = 10\mu$m. 
  }
\label{fig:theta}
  \end{center}
\end{figure}

It is clear from these results that the model is able to predict the steady-state
space-charge affected apex field $E_P$ and the
injected current $I_{\inj}$ reasonably well, when the space-charge effect is moderate.
We can use the values obtained from the model
to quantify the severity of space-charge. Fig.~\ref{fig:theta} shows
a plot of $\vartheta$ as a function of $E_L$ for the model considered where
$D = 10\mu$m and the apex enhancement factor is $\gamma_a = 4.715$. Note that the upper limit of
the vacuum field used in the Fig.~\ref{fig:theta} and \ref{fig:FN}
is $E_L \simeq 13.2$V/nm. This is to ensure that the top of the potential barrier remains
above the Fermi level. The values of $\vartheta$ for $E_L$ exceeding 10V/nm
or $\vartheta < 0.8$ are not very  accurate due to larger deviations from the cosine law.
They are however expected to be indicative of the general trend.

\begin{figure}[hbt]
  \begin{center}
    \vskip -0.75cm
\hspace*{-.5cm}\includegraphics[scale=0.34,angle=0]{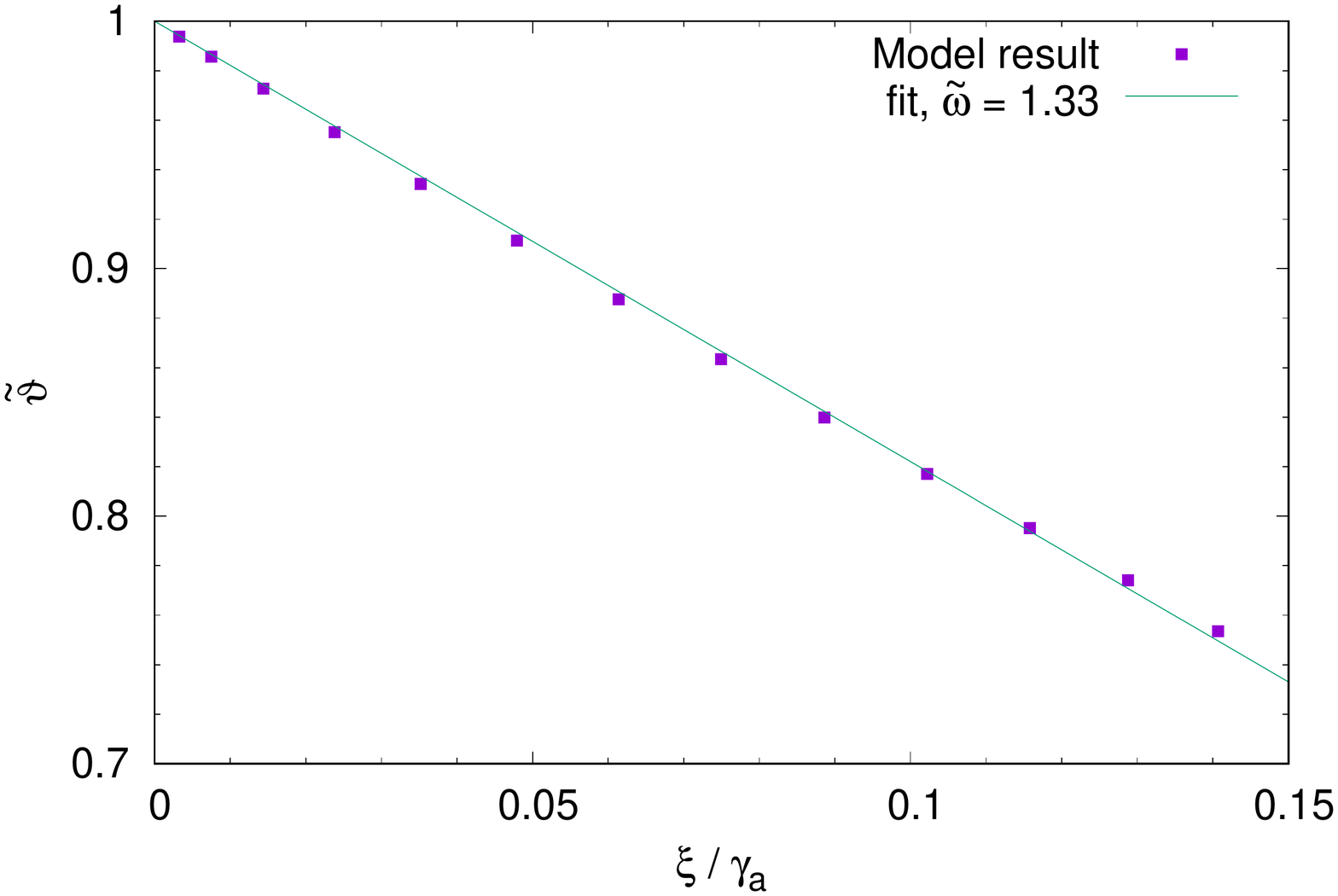}
\vskip -0.6cm
\caption{The normalized scaled space charge affected field $\tilde{\theta}$ is plotted
  against the normalized space charge affected current ${\xi}$.
  It follows approximately the straight line Eq.~(\ref{eq:small1}) with
  the parameter $\tilde{\omega}  \approx 1.33$.
  }
\label{fig:xi-thet}
  \end{center}
\end{figure}

The model results can also be cast in a (${\xi},\tilde{\theta}$) plot for completeness.
This is shown in Fig.~\ref{fig:xi-thet} for $D = 10\mu$m. Also shown is the straight
line of Eq.~(\ref{eq:small1}) with the parameter $\tilde{\omega} \approx 1.33$, obtained
by a least square fit.

\begin{figure}[hbt]
  \begin{center}
    \vskip -0.75cm
\hspace*{-.5cm}\includegraphics[scale=0.34,angle=0]{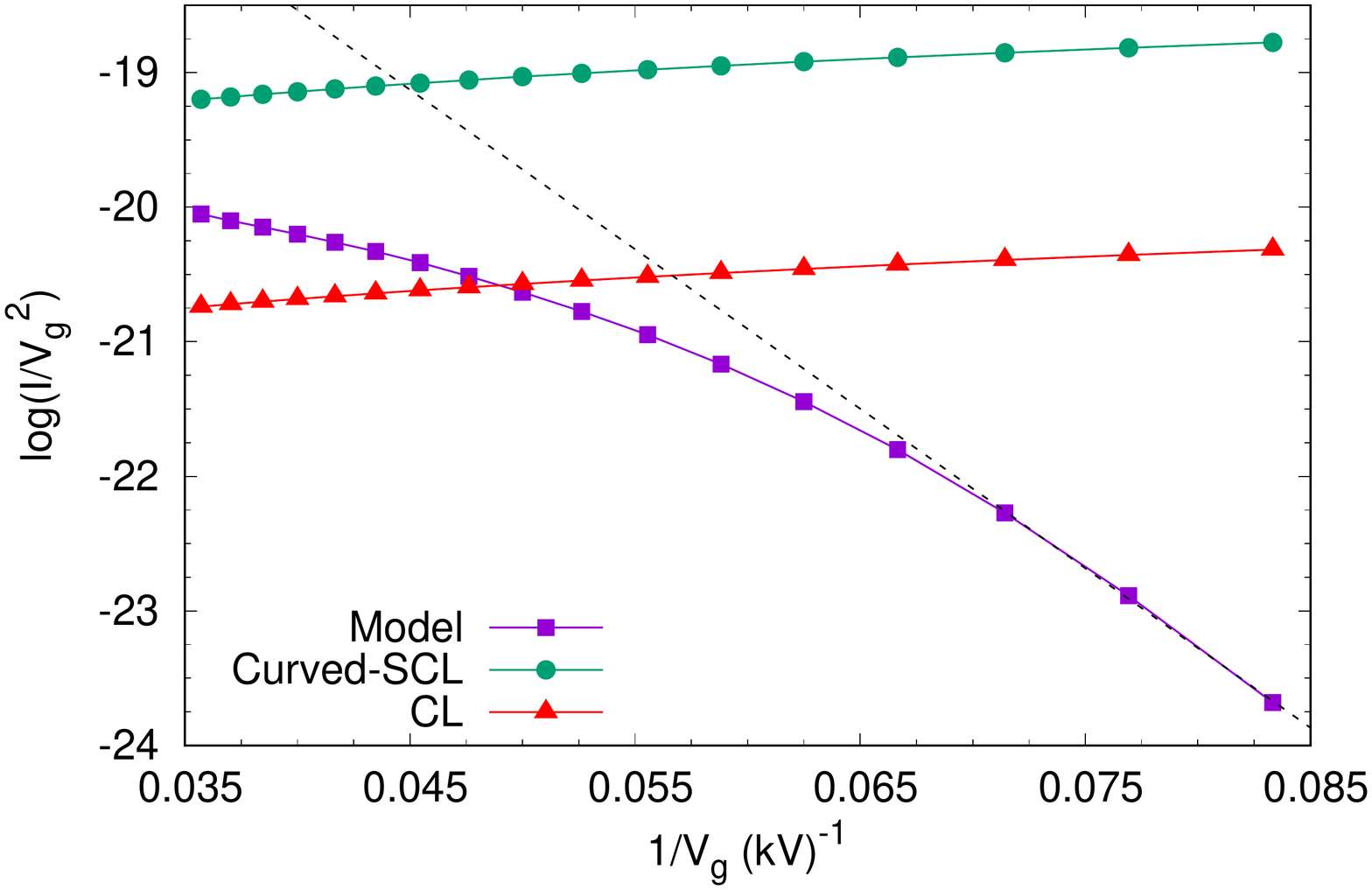}
\vskip -0.6cm
\caption{The steady state field emission current $I = I_{\inj}$ predicted by the model is
  shown as an FN-plot against the applied diode voltage $V_g$. The dashed line shows
  the straight line fit at lower voltages. Also show are (i) the FN-plot of the Child-Langmuir current
  $I = \pi b^2 J_{\text{\CL}}$ and the space charge limited current from a curved emitter $I = I_{\text{Curved-SCL}}$.
  The field emission current is greater than planar space charge limited current from the
  base of the curved emitter for $E_0 > 2$V/nm. It is however less than the
  space charge limited current from a curved emitter. 
  }
\label{fig:FN}
  \end{center}
\end{figure}

Finally, the data obtained using the space-charge affected field emission model
is shown as an FN plot in Fig.~\ref{fig:FN}. Clearly,
the space charge affected field emission current deviates from the straight line
as observed earlier \cite{barbour53,jensen97a,jensen97b,forbes2008}.
Also shown alongside are FN plots of 2 different space charge limited currents.
The first of these (solid triangles in Fig.~\ref{fig:FN})  is the planar Child-Langmuir current from an area equal
to the base of the curved emitter. Thus, $I = J_{\CL} \pi b^2$ where

\be
J_{\text{\CL}} = \frac{4}{9} \left( \frac{2e}{m} \right)^{1/2} \frac{V_g^{3/2}}{D^2}  \label{eq:CL}
\ee

\noi
is the space charge limited current from a planar surface.
The second (solid circles in Fig.~\ref{fig:FN}) uses the recently reported 
approximate space charge limited current for curved emitters

\be
I_{\text{Curved-SCL}} \simeq \gamma_a \pi b^2 \gamma_a J_{\CL} \label{eq:SCL}
\ee

\noi
which reduces to the planar case for $\gamma_a = 1$.

The space-charge affected field emission current can thus exceed the planar space
charge limit but is bounded by the space charge limited current for a curved emitter.

\section{Discussions}

We have put forward a time-dependent model for space-charge affected field emission that
is simple to implement and takes only a few seconds to yield the time evolution of the
apex field and emitted current over several transit times. The steady state values achieved
were compared with the PIC code PASUPAT and found to be in good agreement
keeping the only free parameter $\alpha$ fixed at $1/4$ which corresponds to emission from
a circular patch. The agreement was excellent under low and moderate space-charge conditions
corresponding to smaller transit times or lower vacuum fields, for which the agreement
with the cosine law of field variation is good.

The model is also applicable to a cluster of emitters or a large area field emitter (LAFE)
arranged randomly or in an ordered fashion
so long as the individual field enhancement factors
are known\cite{db_rudra1,rudra_db,db_rudra2,anodeprox,db_hybrid} and
two emitters are not too close to each other for mutual space charge
effects to kick in. In the example chosen to verify the model
is that of emitters on an infinite square lattice  with Neumann boundary condition
on the transverse computational boundaries placed at half the lattice constant.
The effect of other emitters is manifested in the slightly lower field enhancement
factor due to the shielding effect. In general, if $h/R_a$ is large enough ($ > 25$),
the hybrid model\cite{db_rudra2,db_hybrid} can be used to calculate the field enhancement factor of individual
emitters in a LAFE and hence the net space-charge affected field emission current
may be calculated. 

\section{Acknowledgements} PASUPAT simulations were performed on ANUPAM-AGANYA super-computing facility at Computer Division, BARC.

\section{Author Declarations}

\subsection{Conflict of interest} The authors have no conflicts to disclose.
\subsection{Data Availability} The data that supports the findings of this study are available within the article.

\vskip 0.25 in
{\it Data Availability}: The data that supports the findings of this study are available within the article.\\

\vskip -0.75 in
\section{References} 
\vskip -0.2in


\begin{thebibliography}{99}
\bibitem{alternate} An alternate measure of space charge strength is the ratio of
  the steady-state field emission current and the space charge limited current. Note however
  that the SCL current depends linearly on the apex field enhancement factor \cite{db_scl}.
\end{thebibliography}
\end{document}